\documentstyle[psfig,12pt]{article}
\def\TL{\hfil$\displaystyle{##}$}
\def\TR{$\displaystyle{{}##}$\hfil}
\def\TC{\hfil$\displaystyle{##}$\hfil}
\def\TT{\hbox{##}}
\def\seqalign#1#2{\vcenter{\openup1\jot
  \halign{\strut #1\cr #2 \cr}}}

\def\comment#1{}
\def\fixit#1{}

\def\tf#1#2{{\textstyle{#1 \over #2}}}

\def\mop#1{\mathop{\rm #1}\nolimits}

\def\Vol{\mop{Vol}}
\def\vol{\mop{vol}}

\def\tr{\mop{tr}}

\def\lsim{\mathrel{\mathstrut\smash{\ooalign{\raise2.5pt\hbox{$<$}\cr\lower2.5pt\hbox{$\sim$}}}}}
\def\gsim{\mathrel{\mathstrut\smash{\ooalign{\raise2.5pt\hbox{$>$}\cr\lower2.5pt\hbox{$\sim$}}}}}
\def\sqr#1#2{{\vcenter{\vbox{\hrule height.#2pt
         \hbox{\vrule width.#2pt height#1pt \kern#1pt
            \vrule width.#2pt}
         \hrule height.#2pt}}}}
\def\square{\mathop{\mathchoice\sqr56\sqr56\sqr{3.75}4\sqr34\,}\nolimits}
\def\href#1#2{#2}  

\def\lbldef#1#2{\expandafter\gdef\csname #1\endcsname {#2}}
\def\eqn#1#2{\lbldef{#1}{(\ref{#1})}%
\begin{equation} #2 \label{#1} \end{equation}}
\def\eqalign#1{\vcenter{\openup1\jot
    \halign{\strut\span\TL & \span\TR\cr #1 \cr
   }}}

\def\arccoth{\mop{arccoth}}
\begin{document}
\baselineskip=15.5pt
\pagestyle{plain}
\setcounter{page}{1}
\renewcommand{\thefootnote}{\fnsymbol{footnote}}
%--------+---------+---------+---------+---------+---------+---------+
%Title page
\begin{titlepage}

\begin{flushright}
HUTP-{\sl A}010\\
hep-th/9902155
\end{flushright}
\vfil

\begin{center}
{\huge Dilaton-driven confinement}
\end{center}

\vfil
\begin{center}
{\large Steven S. Gubser}\\
\vspace{1mm}
Lyman Laboratory of Physics \\
Cambridge, MA  02138 \\
Tel: (617) 496-8362  \\
Fax: (617) 496-8396 \\
ssgubser@born.harvard.edu \\
\end{center}

\vfil

\begin{center}
{\large Abstract}
\end{center}

\noindent
 We derive a solution of type~IIB supergravity which is asymptotic to
$AdS_5 \times S^5$, has $SO(6)$ symmetry, and exhibits some of the features
expected of geometries dual to confining gauge theories.  At the linearized
level, the solution differs from pure $AdS_5 \times S^5$ only by a dilaton
profile.  It has a naked singularity in the interior.  Wilson loops follow
area law behavior, and there is a mass gap.  We suggest a field theory
interpretation in which all matter fields of ${\cal N}=4$ gauge theory
acquire a mass and the infrared theory is confining.

\vfil
\begin{flushleft}
February 1999
\end{flushleft}
\end{titlepage}
\newpage
%--------+---------+---------+---------+---------+---------+---------+
%Body
\section{Introduction}
\label{Introduction}

One of the interesting aspects of the AdS/CFT conjecture
\cite{juanAdS,gkPol,witHolOne} is that energy plays the role of an extra
dimension: a fifth dimension if we are working with four-dimensional
quantum field theories.  The expectation is that the gravitational
equations which dictate the bulk geometry to be equivalent in some sense to
the renormalization group (RG) equations for the quantum field theory
(QFT).  It is difficult from a mathematical point of view to see how this
can be, since the RG equations are first order while the supergravity
equation are second order.  It is also difficult to gain intuition
regarding the problem from examining the conformal case, since there the
anti-de Sitter is self-similar with respect to motions in the fifth
dimension, and the evolution under RG is trivial.

It is desirable, then, to come up with examples where the quantum field
theory is not conformal.  The obvious approach is to start with ${\cal
N}=4$ super-Yang-Mills theory and add some relevant operators,
corresponding to tachyon fields ($m^2 < 0$) in $AdS_5$.  It turns out to be
very difficult to find a supergravity solution with such fields excited
because they all have $SO(6)$ quantum numbers: that is, they come from
ten-dimensional fields with some non-trivial variation on $S^5$.  Recently,
progress has been made in this direction by considering the truncation of
the Kaluza-Klein (KK) reduction of type~IIB supergravity on $S^5$ to
five-dimensional gauged supergravity \cite{ppz,dz}.  This amounts to
keeping only the first few spherical harmonics for each ten-dimensional
field in the five-dimensional theory.  In this setup, RG flows in a
boundary QFT are supposed to be reflected in a dependence on the fifth
dimension of the scalars in the gauged supergravity theory.  It has been
speculated \cite{kpw} that any solution of five-dimensional maximally
supersymmetric gauged supergravity can be lifted to an exact solution of
the ten-dimensional theory, in analogy with the embedding of maximally
supersymmetric four-dimensional gauged supergravity in eleven-dimensional
supergravity \cite{deWitN}.  So far, the only examples in which the flow
equations are known explicitly break all supersymmetry.

In this paper, we wish to take an approach which is much simpler from the
point of view of supergravity calculations: we will study geometries which
preserve the full $SO(6)$ invariance as well as the 3+1-dimensional
Poincar\'e symmetry of the boundary quantum field theory.  It would seem to
follow from the no-hair theorems that any such geometry other than the full
D3-brane metric must have a naked singularity, and this indeed is the case
for the geometry which we will exhibit in section~\ref{Solution}.  Unlike
the full D3-brane metric, our geometry is asymptotic to $AdS_5 \times S^5$
far from the singularity.  It is tempting to guess that its field theory
``image'' (in the holographic sense of AdS/CFT) is ${\cal N}=4$
super-Yang-Mills theory deformed by a relevant operator.  The difficulty is
that at strong coupling, AdS/CFT itself predicts that there are no relevant
operators which are $SO(6)$ singlets.  In section~\ref{FieldTheory} we will
suggest that a solution to this dilemma is to add to the lagrangian an
$SO(6)$-invariant mass term for scalars.  This operator corresponds to an
excited string state in $AdS_5 \times S^5$.

In section~\ref{Confine} we show that our geometry satisfies the usual
criterion for confinement, namely the area law for Wilson loops and a mass
gap.  This is hardly unexpected from the perturbative field theory point of
view, since confinement is the generic behavior for gauge theory coupled to
massive matter.  Finally, in section~\ref{Discuss}, we discuss the global
structure of the geometry and obtain the conditions under which
supergravity is a valid approximation.

\section{The equations}
\label{Equations}

In ten-dimensional type~IIB supergravity, the most general ansatz with
$SO(6)$ symmetry, 3+1-dimensional Poincar\'e invariance, and $N$ units of
five-form flux through an $S^5$ is
  \eqn{SAnsatz}{\eqalign{
   ds_{10}^2 &= \hat{g}_{MN} dx^M dx^N 
    = e^{-{10 \over 3} \chi + 2\sigma} (-dt^2 + dx_1^2 + dx_2^2 + dx_3^2 + 
      dz^2) + L^2 e^{2\chi} d\Omega_5^2  \cr
   F_5 &= {N \sqrt{\pi} \over 2 \Vol S^5}
     \left( \vol_{S^5} + * \vol_{S^5} \right) \ ,
  }}
 where $\chi$, $\sigma$, and also the dilaton $\phi$ and the axion $C$, are
allowed to depend only on the radial coordinate $z$.  The dilaton and the
axion combine to form the complex coupling $\tau = C + i e^{-\phi}$, which
in the AdS/CFT correspondence is identified with $\theta/2\pi + 4\pi
i/g_{YM}^2$.  We will always work in Einstein frame unless otherwise noted
explicitly.

The relevant equations of type IIB supergravity, truncated to the fields we
are interested in, read \cite{GSW}
  \eqn{RelvantIIB}{\eqalign{
   \hat{\square} \phi &= e^{2\phi} (\partial_M C)^2  \cr
   \hat{\square} C &= -2 (\partial_M \phi) (\partial^M C)  \cr
   \hat{R}_{MN} &= \tf{1}{2} \partial_M \phi \partial_N \phi + 
    \tf{1}{2} e^{2\phi} \partial_M C \partial_N C + 
    {\kappa^2 \over 6} F_{M P_1 \ldots P_4} F_N{}^{P_1 \ldots P_4} \ .
  }}
 The Einstein equations in the $S^5$ directions are satisfied if
  \eqn{LNRel}{
   L^4 = {\kappa N \over 2 \pi^{5/2}} \ .
  }
 The remaining equations can be expressed in purely five-dimensional terms:
  \eqn{Feoms}{\eqalign{
   \square \phi &= e^{2\phi} (\partial_\mu C)^2  \cr
   \square C &= -2 (\partial_\mu \phi) (\partial^\mu C)  \cr
   \square \chi &= {4 \over L^2} \left( e^{-{16 \over 3} \chi} - 
    e^{-{40 \over 3} \chi} \right)  \cr
   R_{\mu\nu} &= \tf{1}{2} \partial_\mu \phi \partial_\nu \phi + 
    \tf{1}{2} e^{2\phi} \partial_\mu C \partial_\nu C + 
    \tf{40}{3} \partial_\mu \chi \partial_\nu \chi - 
    {g_{\mu\nu} \over L^2} \left( \tf{20}{3} e^{-{16 \over 3} \chi} - 
     \tf{8}{3} e^{-{40 \over 3} \chi} \right) \ ,
  }}
 where now the metric
  \eqn{FiveMet}{
   ds_5^2 = g_{\mu\nu} dx^\mu dx^\nu = 
    e^{2\sigma} (-dt^2 + dx_1^2 + dx_2^2 + dx_3^2 + dz^2)
  }
 is used to compute $R_{\mu\nu}$ and to contract indices.  The equations
\Feoms\ follow from the five-dimensional action
  \eqn{FAction}{
   S = {1 \over 2 \kappa_5^2} 
    \int d^5 x \sqrt{g} \left[ R - \tf{1}{2} (\partial_\mu \phi)^2 - 
     \tf{1}{2} e^{2\phi} (\partial_\mu C)^2 - 
     \tf{40}{3} (\partial_\mu \chi)^2 + {1 \over L^2} 
      \left( 20 e^{-{16\over 3} \chi} - 8 e^{-{40\over 3} \chi} \right)
      \right] \ ,
  }
 as was first noted in \cite{gktRun}.  The gravitational couplings
$\kappa_5$ and $\kappa$ in five and ten dimensions are related by
  \eqn{FKappa}{
   {1 \over 2 \kappa_5^2} = {\pi^3 L^5 \over 2 \kappa^2} = 
    {N^2 \over 8 \pi^2} {1 \over L^3} \ .
  }
 The Weyl factor on the non-compact part of the metric in \SAnsatz\
was chosen expressly so that the wave equation $\hat{\square} \eta =
0$ for a minimal scalar in the $s$-wave would reduce to the
five-dimensional wave equation $\square \eta = 0$.  This is the same
requirement that makes $ds_5^2$ the Einstein frame metric.

It is possible to reduce the equations of motion \FAction\ to a set of
coupled non-linear second order ordinary differential equations in $\phi$,
$C$, $\chi$, and $\sigma$.  These equations seem too complicated to deal
with in general, but there is an obvious simplification: $C=0$ and
$\chi=0$.  Then the equations become much simpler:
  \eqn{OtherEin}{\eqalign{
   \partial_z^2 \sigma + 3 (\partial_z \sigma)^2 &= 
     {4 \over L^2} e^{2\sigma}  \cr
   4 \partial_z^2 \sigma &= -\tf{1}{2} (\partial_z \phi)^2 + 
     {4 \over L^2} e^{2\sigma}  \cr
   e^{-5 \sigma} \partial_z e^{3 \sigma} \partial_z \phi &= 0 \ .
  }}
 In future sections we will usually deal with the full ten-dimensional
Einstein metric.  Because $\chi=0$ there is no distinction between the
five-dimensional Einstein metric and the ten-dimensional Einstein metric
restricted to the five-dimensional non-compact space.  So we will drop the
hats from ten-dimensional quantities.

\section{The solution}
\label{Solution}

\begin{figure}
   \vskip0cm
   \centerline{\psfig{figure=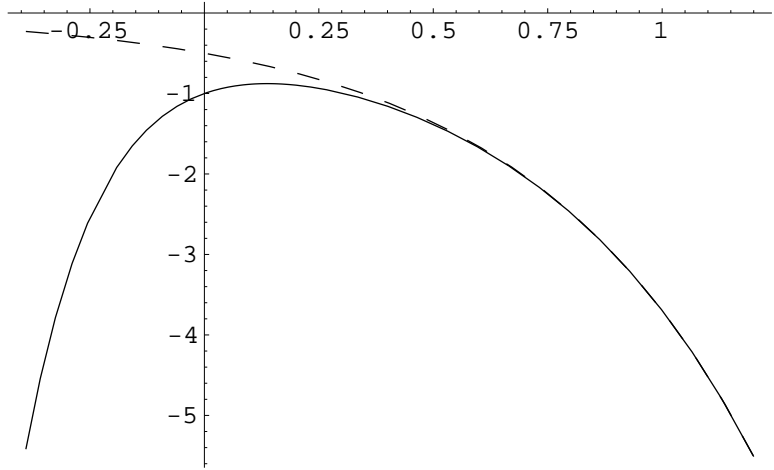,width=4in}}
   \vskip0cm
 \caption{$V(\sigma)$ as a function of $\sigma$ for $B^2=24$ (solid line)
and $B=0$ (dashed line).}\label{figA}
  \end{figure}

The last equation in \OtherEin\ can be integrated directly to give
  \eqn{EliminateDilaton}{
   \phi(z) = \phi_\infty + {B \over L} \int_0^z d\tilde{z} \, 
    e^{-3 \sigma(\tilde{z})} \ ,
  }
 where $\phi_\infty$ is the value of the dilaton at the boundary of the
asymptotically $AdS_5$ geometry, and $B$ is another arbitrary constant.  We
can take $B>0$ since $S$-duality sends $\phi \to -\phi$ while preserving
the Einstein metric.  Substituting \EliminateDilaton\ back into \OtherEin,
defining $u = z/L$, and rearranging slightly, one obtains
  \eqn{TwoForOne}{\eqalign{
   \partial_u^2 \sigma &= e^{2\sigma} - {B^2 \over 8} e^{-6\sigma}  \cr
   (\partial_u \sigma)^2 &= e^{2\sigma} + {B^2 \over 24} e^{-6\sigma} \ .
  }}
 The first equation follows from differentiating the second, so we see that
\OtherEin\ is a consistent system of equations despite being
overdetermined.  Physically, \TwoForOne\ describes the zero-energy
trajectory of a classical particle with unit mass moving in the potential
$V(\sigma) = -\tf{1}{2} e^{2\sigma} - {B^2 \over 48} e^{-6\sigma}$, which
is depicted in figure~\ref{figA}.
 We regard $z = uL$ as a radial variable which goes to $0$ at the boundary
of the geometry, where $\sigma$ becomes large.  Thus in our mechanical
analog, we are starting the particle at $\sigma=\infty$ at ``time'' $u=0$.
In finite time the particle gets to $\sigma = 0$.  If $B=0$, then it
continues to move to the left in figure~\ref{figA} ever more slowly as
$u\to\infty$.  But for $B \neq 0$, the particle reaches a minimum velocity
at some $\sigma < 0$, and then falls back down the other side of the
potential after a finite time $u_0$.  The $B=0$ solution is pure $AdS_5$ with
constant dilaton.  Not unexpectedly, the $B\neq 0$ geometry is geodesically
incomplete and singular at the point $u=u_0$.  To find $u_0$ explicitly, we
use the standard trick for integrating the motion in a one-dimensional
potential well:
  \eqn{OneDimInt}{\eqalign{
   u &= \int_\sigma^\infty {d\tilde\sigma \over \sqrt{-2 V(\tilde\sigma)}}
     = \int_\sigma^\infty {d\tilde\sigma \over 
        \sqrt{e^{2\tilde\sigma} + {B^2 \over 24} e^{-6\tilde\sigma}}}  \cr
     &= {3^{1/8} \Gamma(3/8) \Gamma(1/8) \over 8^{7/8} \sqrt{\pi} B^{1/4}} -
        \sqrt{{8 \over 3}} {e^{3\sigma} \over B} 
        F\left( {3 \over 8},{1 \over 2};{11 \over 8};
         -{24 e^{8 \sigma} \over B^2} \right)
  }}
 where $F(\alpha,\beta;\gamma;z)$ is the usual hypergeometric function.
The second term vanishes as $\sigma \to -\infty$, so we find $u_0 =
{3^{1/8} \Gamma(3/8) \Gamma(1/8) \over 8^{7/8} \sqrt{\pi} B^{1/4}}$.

We can find the dilaton explicitly in terms of $\sigma$ by performing the
integral in \EliminateDilaton:
  \eqn{GotDilaton}{
   \phi = \phi_\infty + \sqrt{3 \over 2} 
    \arccoth \sqrt{1 + {24 \over B^2} e^{8\sigma}} \ .
  }
 Note that $\phi$ increases as $\sigma$ decreases, and $\phi\to \infty$ as
$\sigma \to -\infty$, which is what we expect since the far interior of the
geometry ($\sigma \to -\infty$) corresponds to the infrared in the gauge
theory.  We can also write the ten-dimensional Einstein metric explicitly
if we use $\sigma$ rather than $z$ as the radial variable:
  \eqn{ExplicitMetric}{
   ds_{10}^2 = e^{2\sigma} (-dt^2 + dx_1^2 + dx_2^2 + dx_3^2) + 
    {L^2 d\sigma^2 \over 1 + {B^2 \over 24} e^{-8\sigma}} + 
    L^2 d\Omega_5^2 \ .
  }
 From \GotDilaton\ and \ExplicitMetric\ it is evident that we can cancel
the factors of ${B^2 \over 24}$ by sending $\sigma \to \sigma + {1 \over 8}
\log {B^2 \over 24}$.  In order not to reintroduce factors of ${B^2 \over
24}$ in the world-volume components of the metric, we should also rescale
$t \to \left( {B^2 \over 24} \right)^{-1/8} t$ and $x_i \to \left( {B^2
\over 24} \right)^{-1/8} x_i$ for $i=1,2,3$.  If we also rescale $z \to
\left( {B^2 \over 24} \right)^{-1/8} z$ then the net result is the same as
if we had set $B^2 = 24$ throughout this section.  Choice of the radial
coordinate in $AdS_5$ corresponds to choice of one out of a given class of
conformally equivalent boundary metrics.  Thus we see that the freedom to
change $B$ in the solution \GotDilaton, \ExplicitMetric\ corresponds merely
to the asymptotic scale invariance of the boundary theory.

\section{Confinement}
\label{Confine}

\begin{figure}
   \vskip0cm
   \centerline{\psfig{figure=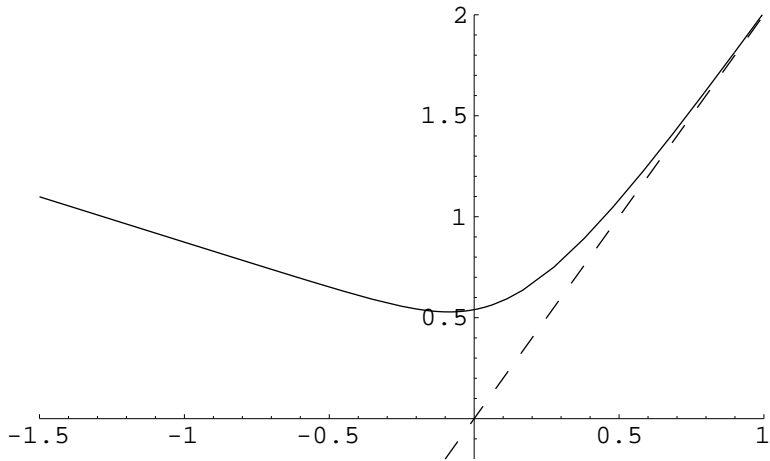,width=4in}}
   \vskip0cm
 \caption{$\log n = 2\sigma + (\phi-\phi_\infty)/2$ as a function of
$\sigma$ for $B^2=24$ (solid line) and $B=0$ (dashed line).}\label{figB}
  \end{figure}

The string coupling gets strong as $\sigma \to -\infty$, but this does not
necessarily mean confinement, as illustrated in \cite{ktZeroTwo}, where the
far interior of a D3-brane geometry in a type 0 string theory was found to
be $AdS_5 \times S^5$ geometry with formally infinite dilaton.\footnote{In
\cite{minConfine}, more generic interior geometries were explored and shown
to exhibit properties of confinement.  These solutions are similar in
spirit and in global structure to ours.  We thank J.~Minahan for bringing
this work to our attention.}  The most straightforward test of confinement
in the context of holography \cite{witHolTwo} is the area law for Wilson
loops.  The metric felt by strings is not the Einstein metric
\ExplicitMetric, but rather the string metric
  \eqn{StringMetricForm}{\eqalign{
   ds_{\rm string}^2 &= G_{MN} dx^M dx^N 
      = e^{(\phi-\phi_\infty)/2} ds_{\rm Einstein}^2  \cr
     &= e^{2\sigma + (\phi-\phi_\infty)/2}
        \left( -dt^2 + dx_1^2 + dx_2^2 + dx_3^2 + dz^2 \right) + 
       L^2 e^{(\phi-\phi_\infty)/2} d\Omega_5^2 \ .
  }}
 Following \cite{juanWilson,ReyYee,witHolTwo}, we consider a fundamental
string following a trajectory ending on specified points of the boundary of
the asymptotically $AdS_5$ geometry which minimizes the Nambu-Goto action,
  \eqn{NambuGoto}{\eqalign{
   S_{NG} &= {1 \over 2\pi\alpha'} \int d^2\xi \, 
    \sqrt{\det \left( G_{MN} {\partial x^M \over \partial\xi^\alpha}
      {\partial x^N \over \partial\xi^\beta} \right)}  \cr
     &= {1 \over 2\pi\alpha'} \int dt dx \,
    e^{2\sigma + (\phi-\phi_\infty)/2} 
     \sqrt{1 + \left( {dz \over dx} \right)^2} \ .
  }}
 In the second line we have assumed that the string is at a constant
angular position in $S^5$, and we have set $\xi^0 = t$, $\xi^1 = x_1 = x$,
and $x_2 = x_3 = 0$.  The time integral is trivial, so we see that the
integral whose minimum determines the string trajectory is
  \eqn{FermatForm}{
   V = {1 \over 2\pi\alpha'} \int dx \, 
    n(z) \sqrt{1 + \left( {dz \over dx} \right)^2} \ ,
  }
 where we have defined $n(z) = e^{(\phi(z)-\phi_\infty)/2 + 2\sigma(z)}$.
The form \FermatForm\ is suggestive of Fermat's Principle for the path of
light rays: if $n(z)$ is regarded as the ``refractive index,'' then $V$ is the
total ``time'' (or more properly, the total distance in the flat metric
$ds^2 = dx^2 + dz^2$) it takes to traverse the trajectory followed by the
``light ray'' (more properly, the string).  As one can see from
figure~\ref{figB}, $n(z)$ has a global minimum when $B \neq 0$, namely $n_*
= (\sqrt{2} + \sqrt{3})^{\sqrt{3/8}} \left( {B^2 \over 48} \right)^{1/4}$
at $\sigma_* = {1 \over 8} \log {B^2 \over 48}$.
 In the light ray analogy, confinement is realized as total internal
reflection: for large separation of endpoints, the trajectory which
minimizes $V$ locates itself very nearly at the minimum of $n(z)$ for most
of its length.  This is similar to the way fiber optics work: the light ray
is guided to the minimum of $n(z)$.  When the separation $r$ of the two
endpoints is much larger than $L$, we have
  \eqn{LongRangeV}{
   V(r) \approx {n_* \over 2\pi\alpha'} r \ .
  }
 The relations
  \eqn{GGRels}{\seqalign{\span\TC}{
   L^4 = {N\kappa \over 2\pi^{5/2}} \qquad
   g_{YM}^2 = 4\pi g_s \qquad
   e^{\phi_\infty} = g_s  \cr
   2\kappa^2 = 16\pi G_N = (2\pi)^7 g_s^2 \alpha'^4 
  }}
 result in the simple formula $L^2/\alpha' = \sqrt{g_{YM}^2 N}$.  Using
this in \LongRangeV\ we obtain the QCD string tension as
  \eqn{TQCD}{
   T_{QCD} = {n_* \over 2\pi\alpha'} 
    = {n_* \over 2\pi L^2} \sqrt{g_{YM}^2 N}
    = \left( {B^2 \over 48} \right)^{1/4} (\sqrt{2} + \sqrt{3})^{\sqrt{3/8}}
     {\sqrt{g_{YM}^2 N} \over 2\pi L^2} \ .
  }
 Note that in \GGRels\ we have defined $g_s$ as the string coupling at the
boundary of the asymptotically $AdS_5$ geometry.

The second test of confinement used in \cite{witHolTwo} is the presence of
a mass gap.  This is straightforward to check for perturbations of the
dilaton in our geometry.  Because the dilaton equation of motion $\square
\phi = 0$ is linear, we can ignore the background dilaton profile and
compute the spectrum of $m^2$ in the differential equation
  \eqn{DilGlue}{
   \left[ \sqrt{e^{-8\sigma} + {B^2 \over 24} e^{-16\sigma}} \,
    \partial_\sigma \sqrt{e^{8\sigma} + {B^2 \over 24}} \, \partial_\sigma +
    e^{-2\sigma} m^2 L^2 - \ell(\ell+4) \right] R(\sigma) = 0 \ .
  }
 This radial equation follows from plugging the ansatz $\delta \phi = e^{i
k \cdot x} R(\sigma) Y_\ell$ into the equation $\square\delta\phi = 0$,
where $k^2 = -m^2$ and $Y_\ell$ is a spherical harmonic on $S^5$ with
$\square_{S^5} Y_\ell = \ell (\ell+4) Y_\ell$.  Note that the Einstein
metric, not the string metric, enters into these computations.  A
convenient change of variables is
  \eqn{ChangeVariables}{
   y = \sqrt{2 \over 3} (\phi-\phi_\infty) 
     = \arccoth \sqrt{1 + {24 \over B^2} e^{8\sigma}} \qquad
   E = \sqrt[4]{24} {m^2 L^2 \over \sqrt{B}} \ .
  }
 In these variables, the equation \DilGlue\ reads
  \eqn{DilGlueTwo}{
   \left[ 16 \partial_y^2 - {E \over (\sinh y)^{3/2}} - 
    {\ell (\ell+4) \over \sinh^2 y} \right] R = 0 \ .
  }
 The small $y$ region is near the boundary of the asymptotically $AdS_5$
geometry.  Replacing $\sinh y$ by $y$ in \DilGlueTwo\ results in a form of
Bessel's equation.  In the large $y$ region the potential terms vanish, so
$R(y)$ is linear.  Altogether we have the asymptotics
  \eqn{RAsymp}{
   R(y) \to \left\{ 
    \seqalign{\span\TR \qquad & \span\TT}
     {c_1 \sqrt{y} J_{\ell+2}(4\sqrt{E} y^{1/4}) + 
               c_2 \sqrt{y} N_{\ell+2}(4\sqrt{E} y^{1/4}) &
     for $y \to 0$  \cr
      c_3 + c_4 y & for $y \to \infty$.} \right.
  }
 For $\delta\phi$ to be everywhere small, we must choose $c_2 = c_4 = 0$.
These conditions determine a boundary-value problem with a discrete
spectrum for $E$.  It has been proposed \cite{coot} that the
corresponding massive states in the gauge theory have an interpretation of
glueballs.  Let us label the eigen-energies as $E_{n\ell}$, where
$n=1,2,3,\ldots$ labels the ``glueball'' excitation level and $\ell =
0,1,2,3,\ldots$ is its $SO(6)$ principle quantum number.  From
\ChangeVariables\ we extract the masses:
  \eqn{GlueballMass}{
   m^2_{n\ell} = {\sqrt{B} \over \sqrt[4]{24}} {E_{n\ell} \over L^2} \ .
  }
 We tabulate numerical values for the first few glueball states in
table~\ref{TableA}.  
  \begin{table}
    \begin{center}
    \begin{tabular}{r|ccc}
     $m_{n\ell}^2 L^2 $&$ n=1 $&$ n=2 $&$ n=3 $  \\ \hline
     $\ell=0 $&$ 9.30178 $&$ 30.8883 $&$ 64.8142 $  \\
     $\ell=1 $&$ 15.5574 $&$ 43.4014 $&$ 83.5198 $  \\
     $\ell=2 $&$ 25.0644 $&$ 59.1776 $&$ 105.503 $
    \end{tabular}\vskip-0.5cm
    \end{center}
   \caption{Numerical values for glueball masses with $B^2=24$.}\label{TableA}
  \end{table}
 One sees immediately that the same problem that vexed early efforts
\cite{coot,KKcousins} using the near-extremal D4-brane approach to
$\hbox{QCD}_4$ also occurs here: namely, the ``glueballs'' with $SO(6)$
charge have masses on the same order as the $SO(6)$-neutral glueballs.

The lowest glueball mass associated with the dilaton may not be the minimum
energy excitation $M_{\rm gap}$ of the theory, since there are other
excitations of comparable energy associated with the other supergravity
fields.  However $M_{\rm gap}$ should at least be on the order of the
lowest glueball mass in table~\ref{TableA}.  From \GlueballMass\ and \TQCD\
we see that $T_{QCD}/M_{\rm gap}^2 \sim \sqrt{g_{YM}^2 N}$.  This relation
also obtains in other supergravity models of confinement, but to our
knowledge has not been understood simply in field theory.

\section{Field theory interpretation}
\label{FieldTheory}

In the previous section we saw that the geometry \ExplicitMetric\ exhibits
the basic features of confinement: the area law for Wilson loops and a mass
gap.  Given the ultraviolet behavior of the dilaton, $\phi \sim \phi_\infty
+ {B \over 4} {z^4 \over L^4}$, an obvious interpretation in view of
\cite{bkl} is that we have described a state with nonzero $\langle \tr F^2
\rangle$.  In this section we would like to investigate an alternative
interpretation suggested by the confining properties of the infrared
geometry: namely, that the matter fields of ${\cal N}=4$ super-Yang-Mills
theory have been made massive in an $SO(6)$-invariant fashion, breaking all
supersymmetry as well as conformal invariance, and leaving us with only the
gauge fields below the mass scale.

We can give the scalars an $SO(6)$-invariant mass by adding $m_X^2 \tr
\sum_{I=1}^6 X_I^2$ to the lagrangian.  Also we can give an $SO(6)$
symmetric Majorana mass to the fermions, since they transform in a real
representation of the gauge group.\footnote{We thank A.~Grant for pointing
this out.}  In perturbation theory, either type of mass would induce the
other via loops, and the theory in the infrared would be a confining gauge
theory, perhaps coupled to massive matter if some of it is no heavier than
the confinement scale.  Let us focus on the scalar mass, and write ${\cal
O}_K = \tr \sum_I X_I^2$.

Thinking of ${\cal N}=4$ gauge theory in ${\cal N}=1$ language, the
operator ${\cal O}_K$ is the first component of the Konishi superfield,
$\tr \sum_{i=1}^3 \Phi^\dagger_i e^{V} \Phi_i$, where $\Phi_i$ are the
three chiral matter multiplets.  ${\cal O}_K$ is neither a chiral primary
nor the descendent of a chiral primary, so its dimension is not protected
by the superconformal algebra.\footnote{The discussion that follows is
simplified if ${\cal O}_K$ has definite dimension.  This is true of ${\cal
O}_K = \tr \sum_I X_I^2$ at zero gauge coupling.  As we turn on the gauge
coupling, $\tr \sum_I X_I^2$ may mix with other operators.  If so, then we
think of ${\cal O}_K$ as the resulting operator of definite dimension, and
we expect that ${\cal O}_K$ still has some component of $\tr \sum_I
X_I^2$.}  In \cite{gkPol,witHolOne} it was conjectured that all such
operators are dual to excited string states in $AdS_5 \times S^5$, and that
they acquire dimensions on the order $(g_{YM}^2 N)^{1/4}$ at large 't~Hooft
coupling.  In the absence of a completely satisfactory formalism for
quantizing strings in $AdS_5 \times S^5$, what we mean by excited string
states is simply all fundamental string states except the ones which
participate in the ten-dimensional type~IIB supergravity multiplet.  But
the geometry described in section~\ref{Solution} involved only the
supergravity fields, so how can ${\cal O}_K$ have anything to do with it?
Precisely because $\Delta_K$ is large, the influence of the ``field''
$\phi_K$ dual to ${\cal O}_K$ on the geometry should fall off very quickly
as one moves toward the interior of $AdS_5$.\footnote{We put the word
``field'' in quotes because it is not clear that a field theory of excited
string states exists in any meaningful form.  Nevertheless we will use the
notion of the field $\phi_K$ as an intuitive guide.}  The only effect that
$\phi_K$ could have on the geometry which would persist into the interior
is to source a low-dimension $SO(6)$-invariant field.  The only candidates
(see the tables in \cite{krvn}) are the dilaton, the axion, and the overall
volume of $S^5$---that is, the fields involved in the ansatz \SAnsatz.  The
corresponding operators in the gauge theory are, respectively, $\tr F^2$,
$\tr F\tilde{F}$, and ${\rm Str} \, \left[ F_{\mu_1}{}^{\mu_2}
F_{\mu_2}{}^{\mu_3} F_{\mu_3}{}^{\mu_4} F_{\mu_4}{}^{\mu_1} - {1 \over 4}
(F_{\mu_1}{}^{\mu_2} F_{\mu_2}{}^{\mu_1})^2 \right]$, plus contributions
from the gauginos and scalars, and their dimensions are, respectively,
four, four, and eight.  ${\rm Str}$ is the symmetrized trace.  Let us call
these operators ${\cal O}_4$, $\tilde{\cal O}_4$, and ${\cal O}_8$.  We can
exclude $\tilde{\cal O}_4$ from consideration by insisting on CP
invariance.  The operator ${\cal O}_8$ itself is irrelevant, and so is also
expected to have only a brief influence on the geometry as one flows inward
from the boundary.  If we exclude this operator from consideration, then
the conclusion is that the only way in which a perturbation of $\phi_K$
near the boundary $AdS_5$ can influence the geometry far from the boundary
is to produce a gradient for the dilaton.  It seems guaranteed that
$\phi_K$ will source the dilaton, because the integrated three-point
function
  \eqn{XXPhi}{
   \left\langle {\cal O}_K(x) {\cal O}_K(0) \int {\cal O}_4 \right\rangle 
    \sim g_{YM} {\partial \over \partial g_{YM}} 
     \langle {\cal O}_K(x) {\cal O}_K(0) \rangle 
    \sim g_{YM} {\partial \over \partial g_{YM}} {N^2 \over |x|^{2\Delta_K}} \ ,
  }
 does not vanish when $\Delta_K$ depends on the gauge coupling.  And it
does: $\Delta_K \sim (g_{YM}^2 N)^{1/4}$.

The skeptic may now wonder how it is possible to add a highly irrelevant
operator to ${\cal N}=4$ gauge theory at some high energy scale and see
dramatic effects in the infrared.  The optimist would reply that it is an
extreme example of a dangerous irrelevant operator---by which we mean an
operator which, as a perturbation of the ultraviolet theory, has large
dimension, but which becomes marginal or even relevant along the RG flow.
(Usually in discussions of dangerous irrelevant operators one has in mind
perturbing a theory which is gaussian in the UV, but here we are thinking
of deforming strong-coupling ${\cal N}=4$ gauge theory, which is never
gaussian).  In fact, the supergravity geometry deviates significantly from
$AdS_5$ only for $z \gsim L/\sqrt[4]{B}$, so if the $\phi_K$ perturbation
is located at a much smaller $z$, then there is a large ``scaling region''
(in the sense of critical phenomena: many orders of magnitude in energy) in
which the theory is nearly conformal.  We should add that an operator
${\cal O}_K$ with definite dimension and a finite $\tr \sum_I X_I^2$
component is not the only operator one might consider adding.  In the
language of the optimist, any dangerous irrelevant operator will do,
provided it is an $SO(6)$ singlet, and provided one can argue (along the
lines of \XXPhi, for example) that its dual ``field'' in string theory
triggers a dilaton flow.  Our field theory intuition is that only operators
which are in some sense strong coupling analogs of soft mass breakings or
relevant Yukawa couplings are candidates for dangerous irrelevant
operators.  Confinement is generic behavior once superconformal invariance
is lost.  As we flow toward the infrared past the point where the heavy
string field has damped out, the only relevant information for determining
the subsequent flow is the coefficient $B$ in \GotDilaton.  In short, $B$
parametrizes our ignorance of how the superconformal invariance is broken
in the ultraviolet.

If we are willing to make some crude estimates, we can see how the mass gap
depends on the strength of the ${\cal O}_K$ perturbation.  Suppose we cut
off the geometry at $z=\epsilon$ rather than allowing it to go all the way
out to the natural boundary at $z=0$.  Let us define the theory at
$z=\epsilon$ by setting the value of the dilaton, $\phi = \phi_\infty$, and
also a finite value for the excited string ``field,'' $\phi_K = \mu^2$.  We
regard $\phi_K$ as a dimensionless field, so $\mu$ is a dimensionless mass
parameter: $\phi_K = \mu^2$ on the boundary corresponds to having a term in
the lagrangian of the form $m_X^2 \tr \sum_I X_I^2$ where $m_X =
\mu/\epsilon$.  This is to be compared with the near-extremal D4-brane
approach to $\hbox{QCD}_4$ \cite{witHolTwo}, where $m_{\rm fermions} = \pi
T$ and $T$ is the temperature of the 4+1-dimensional theory.  In both cases
the terms in the bare 3+1-dimensional lagrangian which give matter fields
masses have a coefficient which is an energy squared.

If we regard $\phi_K$ as a linear perturbation of $AdS_5 \times S^5$ with
$\phi = \phi_\infty$ everywhere, then
  \eqn{PhiXEq}{\seqalign{\span\TC}{
   (\square + m_K^2) \phi_K = 0  \cr
   \phi_K = \mu^2 \left( {z \over \epsilon} \right)^{4-\Delta_K} \ .
  }}
 This would be our starting point for computing a two-point function of
${\cal O}_K$.  Beyond the linearized approximation there is a coupling of
$\phi_K$ to the dilaton.  Again crudely, we take as the five-dimensional
lagrangian
  \eqn{LagrangeGuess}{
   {\cal L} = -\tf{1}{2} (\partial\phi)^2 - \tf{1}{2} (\partial\phi_K)^2 - 
     \tf{1}{2} m_K^2 \phi_K^2 + 
      {\lambda \over L^2} \phi_K^2 (\phi-\phi_\infty) \ .
  }
 In order for the relation \XXPhi\ between $\left\langle {\cal O}_K {\cal
O}_K \int {\cal O}_4 \right\rangle$ and $\langle {\cal O}_K {\cal O}_K
\rangle$ to hold, we should pick $\lambda \sim \Delta_K$.  If we set the
same boundary conditions $\phi=\phi_\infty$ and $\partial_z \phi = 0$ at
$z=\epsilon$ as obtained in the pure $AdS_5 \times S^5$ geometry, then we
can solve approximately for the dilaton in the region where the geometry is
still nearly $AdS_5$.  We have 
  \eqn{SolveDil}{\eqalign{
   \square\phi &= {\lambda \over L^2} \phi_K^2  \cr
   {1 \over z^3} \partial_z \phi &= \int_\epsilon^z 
    {d\tilde{z} \over \tilde{z}^5} \lambda \phi_K^2  \cr
    &\approx \lambda \mu^4 \int_\epsilon^\infty {d\tilde{z} \over \tilde{z}^5}
     \left( {\tilde{z} \over \epsilon} \right)^{8-2\Delta_K} 
     = {\lambda \over 2\Delta_K - 4} {\mu^4 \over \epsilon^4} \qquad
     \hbox{for $z \gg \epsilon$} \ .
  }}
 Comparing this with the form which follows from \EliminateDilaton\ in a
nearly $AdS_5$ geometry, namely
  \eqn{CompareToB}{
   \phi \approx \phi_\infty + {B \over 4} {z^4 \over L^4} \ ,
  }
 we find ${B \over L^4} \sim {\lambda \over \Delta_K} {\mu^4 \over
\epsilon^4} \sim {\mu^4 \over \epsilon^4}$ provided $\lambda \sim
\Delta_K$.  The mass gap as computed in section~\ref{Confine} is
  \eqn{MassGapB}{
   M_{\rm gap} \sim {B^{1/4} \over L} \sim {\mu \over \epsilon} = 
    m_X \ .
  }
 We have assumed in this analysis that there is a scaling region, where $z
\gg \epsilon$ but the geometry is still nearly $AdS_5$.  Significant
deviations from $AdS_5$ come at $z = L/B^{1/4} \sim \epsilon/\mu \gg
\epsilon$ provided $\mu \ll 1$, so the story is consistent provided that
the scalar mass $m_X = \mu/\epsilon$ which we put into the theory is much
lower than the scale $1/\epsilon$ at which we define it.  In fact, the
explicit cutoff $\epsilon$ can be taken as small as one likes, and the same
bulk physics will result if one keeps $m_X = \mu/\epsilon$ constant,
provided the dilaton is assumed to be flat at $z=\epsilon$.  Thus, somewhat
surprisingly, the operator ${\cal O}_K \sim \sum_I \tr X_I^2$ when added to
the Wilsonian action as a finite perturbation is effectively dimension two.

The result $M_{\rm gap} \sim m_X$ is similar to what is found in the
near-extremal D4-brane approach to $\hbox{QCD}_4$: there $M_{\rm gap} \sim
m_{\rm fermions}$.  It is rather different from the weak coupling result,
where at one loop one expects $M_{\rm gap} \sim \exp\left( -{8\pi^2 \over
b_0 g_{YM}^2} \right) m_{\rm soft}$.  Here $b_0$ is the leading coefficient
in the beta function below the soft breaking scale: if there are no matter
fields below this scale, then $b_0 = {11 \over 3} N$.\footnote{We thank
O.~Aharony for pointing out to us this comparison with the one-loop
analysis.}  $M_{\rm gap} \sim m_X$ renders comprehensible the presence of
``glueballs'' with $SO(6)$ charge with mass comparable to the lowest
neutral glueball mass $M_{\rm gap}$: the masses of the $SO(6)$-charged
matter fields are comparable to the QCD scale, so we should indeed expect
to see $SO(6)$-charged color singlet states at that same scale, just as we
see strange hadrons in the real world with masses on the order of
$\Lambda_{QCD}$.

It may seem that we have dropped the operator ${\cal O}_8$ from
consideration prematurely: {\it a priori} it seems plausible that the dual
field $\chi$ might be the natural candidate for triggering a dilaton
profile.  Inspection of the equations of motion \Feoms\ reveals that this
is not the case.  In five-dimensional Einstein frame (and still with
$C=0$), the dilaton equation is $\square \phi = 0$ regardless of what
$\chi$ is doing.  Thus constant $\phi$ is always a solution.  As one can
verify from the absence of a $\chi\chi\phi$ term in \FAction, the three
point function $\langle {\cal O}_8 {\cal O}_8 {\cal O}_4 \rangle = 0$,
which is certainly consistent with the fact that $\langle {\cal O}_8 {\cal
O}_8 \rangle$ is independent of $g_{YM}$ if ${\cal O}_8$ is properly
normalized.  With $\phi$ constant and $\chi$ non-constant, one could for
example recover the full D3-brane metric.  In \cite{ghMat} it was suggested
that in a particular large~$g_s N$, small $\alpha'$ double scaling limit
\cite{kDil}, the world-volume gauge theory corresponding to the full
D3-brane metric is
  \eqn{NonRenLag}{
   {\cal L} = {\cal O}_4 + L^4 {\cal O}_8 \ ,
  }
 defined in a Wilsonian sense at a cutoff scale $1/L$.  As usual $L$ is the
radius of the $S^5$ far down the throat of the D3-brane.  The second term
in \NonRenLag\ was thought to characterize the deviations of the geometry
from $AdS_5 \times S^5$ for radii $r \gg L$.  In particular it seemed to
match the scaling form of corrections to the absorption cross-section of
minimally coupled scalars such as the dilaton \cite{ghkk}.

\section{Discussion}
\label{Discuss}

\begin{figure}
   \vskip0cm
   \centerline{\psfig{figure=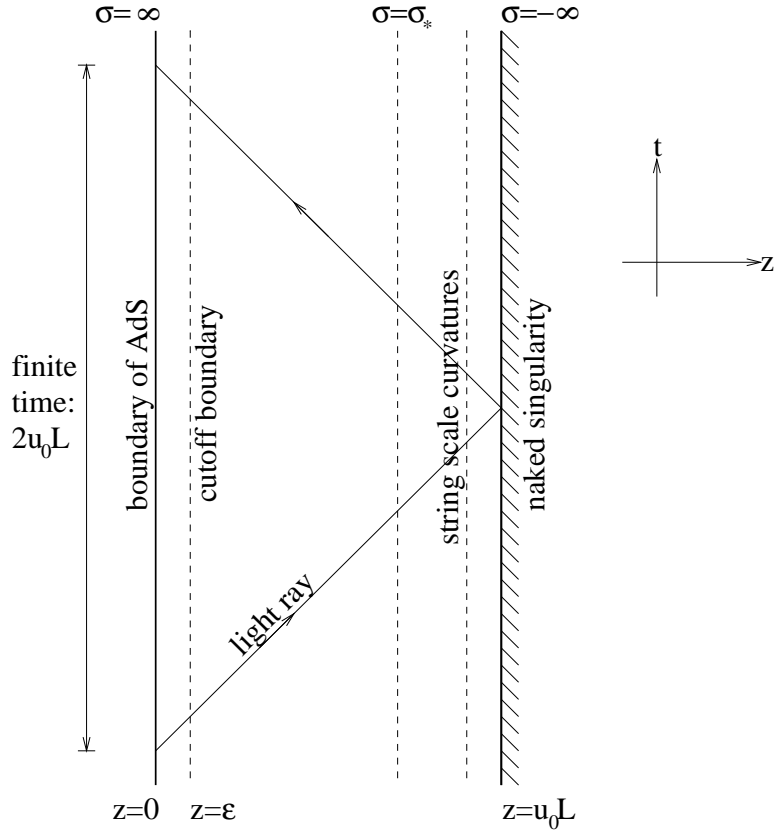,width=4in}}
   \vskip0cm
 \caption{A Carter-Penrose diagram for the solution \ExplicitMetric.  Each
point represents a copy of ${\bf R}^3 \times S^5$.}\label{figC}
  \end{figure}

To summarize our results, it is useful to draw a Carter-Penrose diagram of
the spacetime \ExplicitMetric\ (see figure~\ref{figC}).  
 The conformal structure of the solution \ExplicitMetric\ is most obvious
in the original radial variable $z$.  A light ray sent in from the boundary
can get to the time-like naked singularity and back in finite coordinate
time $\Delta t \sim L/B^{1/4} \sim 1/\Lambda_{QCD}$.

The geometry deviates substantially from anti-de Sitter space for $\sigma
\leq \sigma_*$, where ${B^2 \over 24} e^{-8\sigma_*} = 2$.  The location
$\sigma = \sigma_*$ in the bulk is where long Wilson loops prefer to run,
and the ``refractive index'' $n_* = n(\sigma_*)$ defined in
section~\ref{Confine} is what determines the QCD string tension (that is,
the coefficient on the area law for the Wilson loops).  It is
straightforward to check that the glueball wavefunctions quickly become
flat to the right of this radius in figure~\ref{figC}.

It is interesting to note that the supergravity approximation continues to
be valid far to the right of $\sigma_*$ in figure~\ref{figC}.  Supergravity
is valid if the string coupling $e^{\phi}$ is small and the string frame
curvature is much less than the string scale.  The first condition can be
arranged to hold as close to the singularity at $z=z_0$ as desired,
simply by choosing $g_s = e^{\phi_\infty}$ sufficiently small.  The second
condition means that curvature invariants such as $R$, $R_{MNPQ} R^{MNPQ}$,
etc., computed in string frame, are much smaller than the appropriate power
of $1/\alpha'$.  This will be true if
  \eqn{StringCurvatures}{
   \left( {B^2 \over 24} e^{-8\sigma} \right)^{1 - \sqrt{3/32}} 
     \ll {L^2 \over \alpha'} = \sqrt{g_{YM}^2 N} \ .
  }
 The line labeled ``string scale curvatures'' in figure~\ref{figC}
indicates the radius at which the $\ll$ in \StringCurvatures\ becomes
approximate equality.  The important point is not the peculiar exponent in
\StringCurvatures, but rather the fact that for $N$ and $g_{YM}^2 N$
sufficiently large but $g_{YM}^2$ small, all the physics of glueballs and
Wilson loops takes place in a region of the bulk spacetime where the
supergravity approximation is good.

In section~\ref{FieldTheory}, we proposed to define the theory at a cutoff
$z=\epsilon$, corresponding to an energy $1/\epsilon$, as ${\cal N}=4$
deformed by a $SO(6)$-invariant scalar mass term $m_X^2 \tr \sum_I X_I^2$.
The cutoff served to give the massive bulk scalar field $\phi_K$ dual to
this operator a definite value, $m_X \epsilon$, at $z=\epsilon$.  This
cutoff can be removed, without altering the infrared physics, in the
following way: hold $m_X$ fixed, send $\epsilon$ gradually to zero, and at
each successively smaller value of $\epsilon$ use the boundary conditions
$\phi=\phi_\infty$ and $\partial_z \phi = 0$ at $z=\epsilon$.

The infrared physics we found in section~\ref{Confine} was confinement at a
scale $\Lambda \sim m_X$.  The spectrum of massive excitations did not
appear to be that of pure QCD, nor should it be expected to, since the
adjoint scalars have a mass on the order of the QCD scale.  The adjoint
fermions also get a mass from radiative corrections which is probably also
comparable to $\Lambda$.  Direct comparisons of this supergravity model or
of the others that have been proposed with pure Yang-Mills on the lattice
seem doomed by the presence of ``glueballs'' with charge under a global
symmetry of the massive matter fields.  It would be more relevant, and more
interesting, to compare the supergravity results with some alternative
analysis of a confining gauge theory coupled to adjoint matter fields with
masses on the order of the confinement scale and some global flavor
symmetry.  The vexing Kaluza-Klein glueballs then have a sensible
interpretation as flavored hadrons.

\section*{Note Added}

When this work was near completion, we received \cite{ks}, in which the
solution described in section~\ref{Solution} was independently obtained and
some properties of Wilson loops and the running of the gauge coupling were
discussed in the nearly anti-de Sitter region.

\section*{Acknowledgements}

I would like to thank O.~Aharony, D.~Gross, G.~Horowitz, J.~Minahan,
R.~Myers, J.~Polchinski, N.~Seiberg, A.~Tseytlin, and E.~Witten for useful
discussions.  This research was supported by the Harvard Society of
Fellows, and also in part by the National Science Foundation under grant
number PHY-98-02709, and by DOE grant DE-FGO2-91ER40654.

\newpage

\bibliography{dil}
\bibliographystyle{ssg}

\end{document}